\documentclass{camera}
\usepackage{graphicx}  
\usepackage{epstopdf} 

\begin{document}

%
\title{Competition of $\beta$-delayed protons and $\beta$-delayed $\gamma$ rays in $^{56}$Zn and the exotic $\beta$-delayed $\gamma$-proton decay}

%
\author{S.~E.~A.~Orrigo$^{1,*}$, B.~Rubio$^1$, Y.~Fujita$^2$, B.~Blank$^3$, W.~Gelletly$^4$, J.~Agramunt$^1$, A.~Algora$^{1,5}$, P.~Ascher$^3$, B.~Bilgier$^6$, L.~C{\'a}ceres$^7$, R.~B.~Cakirli$^6$, H.~Fujita$^8$, E.~Ganio{\u{g}}lu$^6$, M.~Gerbaux$^3$, J.~Giovinazzo$^3$, S.~Gr{\'e}vy$^3$, O.~Kamalou$^7$, H.~C.~Kozer$^6$, L.~Kucuk$^6$, T.~Kurtukian-Nieto$^3$, F.~Molina$^{1,9}$, L.~Popescu$^{10}$, A.~M.~Rogers$^{11}$, G.~Susoy$^6$, C.~Stodel$^7$, T.~Suzuki$^8$, A.~Tamii$^8$ \and J.~C.~Thomas$^7$}

%
\organization{$^1$Instituto de F{\'i}sica Corpuscular, CSIC-Universidad de Valencia, E-46071 Valencia, Spain \\
$^2$Department of Physics, Osaka University, Toyonaka, Osaka 560-0043, Japan \\
$^3$Centre d'Etudes Nucl{\'e}aires de Bordeaux Gradignan, CNRS/IN2P3 - Universit{\'e} Bordeaux 1, 33175 Gradignan Cedex, France \\
$^4$Department of Physics, University of Surrey, Guildford GU2 7XH, Surrey, UK \\
$^5$Inst. of Nuclear Research of the Hung. Acad. of Sciences, Debrecen, H-4026, Hungary \\
$^6$Department of Physics, Istanbul University, Istanbul, 34134, Turkey \\
$^7$Grand Acc{\'e}l{\'e}rateur National d'Ions Lourds, BP 55027, F-14076 Caen, France \\
$^8$Research Center for Nuclear Physics, Osaka University, Ibaraki, Osaka 567-0047, Japan \\
$^9$Comisi{\'o}n Chilena de Energ{\'i}a Nuclear, Casilla 188-D, Santiago, Chile \\
$^{10}$SCK.CEN, Boeretang 200, 2400 Mol, Belgium \\
$^{11}$Physics Division, Argonne National Laboratory, Argonne, Illinois 60439, USA \\
$^*$Corresponding author e-mail: sonja.orrigo@ific.uv.es }

\maketitle

\begin{abstract}
Remarkable results have been published recently on the $\beta$ decay of $^{56}$Zn. In particular, the rare and exotic $\beta$-delayed $\gamma$-proton emission has been detected for the first time in the $fp$ shell. Here we focus the discussion on this exotic decay mode and on the observed competition between $\beta$-delayed protons and $\beta$-delayed $\gamma$ rays from the Isobaric Analogue State.
\end{abstract}

%
\section{Introduction}

Decay spectroscopy is a powerful tool for exploring the structure of nuclei at the drip-lines. $\beta$-decay studies, in particular, provide direct access to the absolute values of the Fermi and Gamow-Teller transition strengths, $B$(F) and $B$(GT), respectively.

The proton-rich $^{56}$Zn nucleus was observed for the first time at GANIL in 1999 \cite{Giovinazzo2001}. $^{56}$Zn is a weakly-bound nucleus lying very close to the proton drip-line. It has a quite small proton separation energy, $S_p=$ 560(140) keV \cite{Audi2003}, and third component of the isospin quantum number $T_z$ = -2.

The first study of the $\beta$ decay of $^{56}$Zn was reported in $ref$. \cite{Dossat2007}. More recently, some interesting results on $^{56}$Zn decay have been reported in $ref$. \cite{Orrigo2014}. Among them the discovery of a rare and exotic decay mode, $\beta$-delayed $\gamma$-proton decay, which has been seen for the first time in the $fp$ shell. The consequences of this rare decay sequence for the determination of the Gamow-Teller (GT) strength have also been analyzed.

\section{The experiment}

The experimental study of $^{56}$Zn decay was performed at GANIL in 2010. The experiment used a primary beam of $^{58}$Ni$^{26+}$ to produce $^{56}$Zn. The $^{58}$Ni beam, of 3.7 e$\mu$A and accelerated to 74.5 MeV/nucleon, was fragmented on a natural Ni target, 200 $\mu$m thick. The fragments were selected by the LISE3 separator and implanted into a Double-Sided Silicon Strip Detector (DSSSD). The detection set-up comprised the aforementioned DSSSD detector, 300 $\mu$m thick, a silicon $\Delta E$ detector located 28 cm upstream, and four EXOGAM Ge clovers surrounding the DSSSD.

The EXOGAM clovers were used to detect $\beta$-delayed $\gamma$ rays. The purpose of the DSSSD was the detection of both the implanted fragments and the subsequent charged-particle decays, {\it i.e.}, $\beta$ particles and $\beta$-delayed protons. An implantation event was defined by simultaneous signals in both the $\Delta E$ and DSSSD detectors. A decay event was defined by a signal above threshold (50-90 keV) in the DSSSD and no coincident signal in the $\Delta E$.

The implanted ions were identified and selected by putting a gate in a two-dimensional identification matrix, obtained by combining the energy loss signal from the $\Delta E$ detector and the Time-of-Flight. The latter was defined as the time difference between the cyclotron radio-frequency and $\Delta E$ signal. 

\section{Results on the $\beta$ decay of $^{56}$Zn}

\begin{figure}[!b]
	\centering
	\vspace{-3.0 mm}
	\includegraphics[width=1.\textwidth,clip]{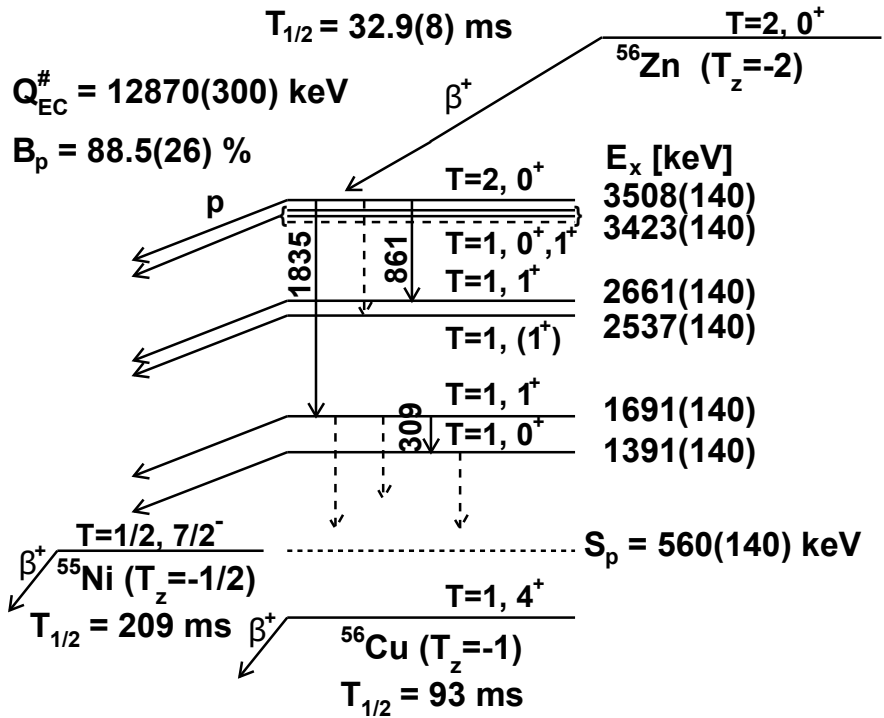}
  \caption{Scheme of the $\beta$ decay of $^{56}$Zn. The solid lines indicate observed proton or $\gamma$ transitions, while the dashed lines correspond to transitions observed in the mirror $^{56}$Co nucleus.}
	\label{decay-scheme}
\end{figure}

The results on the $\beta$ decay of $^{56}$Zn \cite{Orrigo2014} are summarized in the decay scheme in $fig$. \ref{decay-scheme} and in table \ref{table}, and discussed below.

A half-life of $T_{1/2}=$ 32.9(8) ms was obtained for $^{56}$Zn, in agreement with $ref$. \cite{Dossat2007}. To determine $T_{1/2}$, a decay-time spectrum has been constructed from the time correlations between a decay event in a given pixel of the DSSSD (with a total of 256 pixels) and any implantation signal that occurred before and after it in the same pixel, satisfying the identification condition required to select $^{56}$Zn. 

The analysis of the charged-particle spectrum measured in the DSSSD has provided new spectroscopic information on the energy levels populated in the $^{56}$Cu nucleus, the $\beta$-daughter of $^{56}$Zn. These levels are shown in $fig$. \ref{decay-scheme}. The comparison of this level spectrum with that of the mirror $^{56}$Co, obtained by the $^{56}$Fe($^{3}$He,$t$) charge exchange reaction \cite{HFujita2010}, has been very fruitful.

The analysis of the $\gamma$ spectrum measured in the EXOGAM clovers and $\gamma$-proton coincidences have identified three $\gamma$ rays at 309, 861 and 1835 keV.

Absolute $B$(F) and $B$(GT) strengths have been determined (table \ref{table}).

\begin{table}[!h]
	\caption{$\beta$ feedings,  Fermi and Gamow Teller transition strengths to the $^{56}$Cu levels populated in the $\beta^{+}$ decay of $^{56}$Zn.}
	\vspace{3.0 mm}
	\label{table}
	\centering
	  \begin{tabular}{l l l l}
		  \hline
	    $E_X$(keV) & $I_{\beta}$(\%) & $B$(F)  & $B$(GT) \\	\hline
			3508(140)*	 &	  43(5)         &          2.7(5)  &                  \\
		  3423(140)    &    21(1)          &          1.3(5)  & $\leq$0.32\\
		  2661(140)    &    14(1)          &                      & 0.34(6)      \\
		  2537(140)    &   	  0              &                       & 0                 \\
		  1691(140)    &    22(6)          &                      & 0.30(9)       \\
		  1391(140)    &       0              &                      & 0                  \\
			\hline
			*Main component of the IAS.
		\end{tabular}
\end{table}

\subsection{Competition of $\beta$-delayed protons and $\beta$-delayed $\gamma$ rays}

In the first study of the $^{56}$Zn $\beta$ decay \cite{Dossat2007}, the emission of $\beta$-delayed protons was observed but no $\beta$-delayed $\gamma$ rays were seen. This was not a surprise because, in general, in proton-rich nuclei the proton decay is expected to dominate for states well above ($>$1 MeV) the proton separation energy $S_p$. The consequence is that normally the $\beta$ feeding is directly inferred from the measured intensities of the proton peaks. However, cases where there is a competition between $\beta$-delayed proton emission and $\beta$-delayed $\gamma$ de-excitation have also been observed, {\it e.g.}, in $refs$. \cite{Dossat2007,Bhattacharya2008}.

In the $T_{z} = -2 \rightarrow -1, \beta^{+}$ decay of $^{56}$Zn to $^{56}$Cu, the $^{56}$Zn ground state decays with a Fermi transition to its Isobaric Analogue State (IAS) in $^{56}$Cu. It should be noted that the de-excitation of this $T = 2$, $J^\pi = 0^+$ IAS via proton decay to the ground state of $^{55}$Ni ($T = 1/2$, $J^\pi = 7/2^-$) is isospin forbidden. Therefore the proton emission that we observe can only happen through a $T$ = 1 isospin impurity present in the IAS. Moreover in general, when the proton emission is isospin forbidden, the competitive emission of de-exciting $\gamma$ rays from the IAS also becomes possible and can be observed even from IAS lying at an excitation energy well above $S_{p}$ \cite{Dossat2007,Bhattacharya2008}.

The competition between $\beta$-delayed protons and $\gamma$ rays has indeed been observed in $^{56}$Zn. The $\gamma$ decays represent 56(6)\% of the total decays from the 3508 keV IAS. Thus one has to take into account the intensities of both the proton and $\gamma$ peaks to determine the Fermi strength correctly.

We have also found evidence for the fragmentation of $B$(F) due to a strong isospin mixing with a 0$^{+}$ state at 3423 keV \cite{Orrigo2014}, which is important in terms of the mass evaluation \cite{MacCormick2014}. The isospin impurity in the $^{56}$Cu IAS, $\alpha^2$ = 33(10)\% (defined as in $ref$. \cite{HFujita2010}), and the off-diagonal matrix element of the charge-dependent part of the Hamiltonian, $\left\langle{H_c}\right\rangle$ = 40(23) keV, which is responsible for the isospin mixing of the 3508 keV IAS ($T = 2$, $J^\pi = 0^+$) and the 0$^{+}$ part of the 3423 keV level ($T = 1$), are similar to the values obtained in the mirror $^{56}$Co nucleus \cite{HFujita2010}. 

Thus, the proton decay of the IAS proceeds thanks to the $T = 1$ component. However, considering the quite large isospin mixing in  $^{56}$Cu, the much faster proton decay ($t_{1/2}\sim~10^{-18}$ s) should dominate on the $\gamma$ de-excitation ($t_{1/2}\sim~10^{-14}$ s in the mirror). This is not the case since we are still observing the $\gamma$ decay of the IAS in competition with it.

The knowledge on the nuclear structure of the three nuclei involved in the decay, {\it i.e.}, $^{56}$Zn, $^{56}$Cu and $^{55}$Ni, can provide us with a possible explanation for the hindrance of the proton decay. Shell model calculations are in progress to clarify this point.

\subsection{The $\beta$-delayed $\gamma$-proton decay}
 
Besides the competition between $\beta$-delayed proton emission and $\gamma$ decay, the exotic sequence of $\beta$-delayed $\gamma$-proton decay has been detected. Indeed $^{56}$Zn does $\beta$ decay to its IAS in $^{56}$Cu and from there we observe the emission of two $\gamma$ rays of 861 and 1835 keV, populating the $^{56}$Cu levels at 2661 and 1691 keV, respectively. Due to the low $S_p$, these levels are still proton-unbound and thereafter they decay by proton emission. Consequently the rare and exotic $\beta$-delayed $\gamma$-proton decay has been observed. In addition to these two branches, there is a third case. The 1691 keV level emits a $\gamma$ ray of 309 keV, going to the level at 1391 keV that is again proton-unbound and then it de-excites by proton emission.

The $\beta$-delayed $\gamma$-proton decay has been observed here for the first time in the $fp$ shell. This rare decay mode was seen only once before, in the $sd$ shell in $^{32}$Ar \cite{Bhattacharya2008}, but the consequences for the determination of $B$(GT) were not addressed in $ref$ \cite{Bhattacharya2008}. 

The observation of this special decay mode is very important because it does affect the conventional way to determine $B$(GT) near the proton drip-line. For a proper determination of $B$(GT), indeed, it is crucial to correct the intensity of the proton transitions for the amount of indirect feeding coming from the $\gamma$ de-excitation. This finding indicates that it is important to employ $\gamma$ detectors in such studies. This decay mode is expected to be significant in heavier proton-rich nuclei with $T_{z} \leq -3/2$ under study at RIKEN.

\section*{Acknowledgements} 
This work was supported by the Spanish MICINN grants FPA2008-06419-C02-01, FPA2011-24553; Centro de Excelencia Severo Ochoa del IFIC SEV-2014-0398; CPAN Consolider-Ingenio 2010 Programme CSD2007-00042; $Junta~para~la~Ampliaci\acute{o}n~de~Estudios$ Programme (CSIC JAE-Doc contract) co-financed by FSE; MEXT, Japan 18540270 and 22540310; Japan-Spain coll. program of JSPS and CSIC; Istanbul University Scientific Research Projects, Num. 5808; UK Science and Technology Facilities Council (STFC) Grant No. ST/F012012/1; Region of Aquitaine. R.B.C. acknowledges support by the Alexander von Humboldt foundation and the Max-Planck-Partner Group. We acknowledge the EXOGAM collaboration for the use of their clover detectors.

%
\end{document}